\documentstyle[12pt,fullpage]{article}

\newcommand{\be}{\begin{equation}}
\newcommand{\ee}{\end{equation}}
\newcommand{\ba}{\begin{eqnarray}}
\newcommand{\ea}{\end{eqnarray}}
\newcommand{\non}{\nonumber\\ }
\newcommand{\eq}[1]{(\ref{#1})}
\newcommand{\sP}{{\mbox{\bf P}}}
\newcommand{\szb}{{\mbox{\bf z}_2}}
\newcommand{\sza}{{\mbox{\bf z}_1}}
\newcommand{\id}{{I_{16}}}
\newcommand{\cm}{{\mbox{\sl m}}}
\newcommand{\mass}{{\cal M}}
\newcommand{\hP}{{\hat{P}}}
\newcommand{\hX}{{\hat{X}}}
\newcommand{\htheta}{{\hat{{\theta}}}}

\begin{document}

\renewcommand{\thefootnote}{\fnsymbol{footnote}}
\font\csc=cmcsc10 scaled\magstep1
{\baselineskip=14pt
 \rightline{
 \vbox{\hbox{YITP-97-27}
       \hbox{May 1997}
}}}

\vfill
\begin{center}
{\large\bf
BPS Configuration of Supermembrane\\
With Winding in M-direction
}

\vfill

{\csc Kiyoshi EZAWA}\footnote{JSPS fellow}\setcounter{footnote}
{0}\renewcommand{\thefootnote}{\arabic{footnote}}\footnote{
      e-mail address : ezawa@yukawa.kyoto-u.ac.jp},
{\csc Yutaka MATSUO}\footnote{
      e-mail address : yutaka@yukawa.kyoto-u.ac.jp},
{\csc Koichi MURAKAMI}\footnote{
      e-mail address : murakami@yukawa.kyoto-u.ac.jp}\\
\vskip.1in

{\baselineskip=15pt
\vskip.1in
  Yukawa Institute for Theoretical Physics \\
  Kyoto University, Sakyo-ku, Kyoto 606-01, Japan \\
\vskip.1in
}

\end{center}
\vfill

\begin{abstract}
{
We study de Wit-Hoppe-Nicolai supermembrane
with emphasis on the winding in M-direction.
We propose a SUSY algebra of the supermembrane
in the Lorentz invariant form.
We analyze the BPS conditions and argue that the 
area preserving diffeomorphism constraints
associated with the harmonic vector fields 
play an essential role.
We derive the first order partial differential equation
that describes the BPS state with one quarter SUSY.
}
\end{abstract}
\vfill

hep-th/9706002
\setcounter{footnote}{0}
\renewcommand{\thefootnote}{\arabic{footnote}}
\newpage
\vfill

\section{Introduction}


After the struggles to understand the still mysterious
M theory, Matrix theory \cite{r:BFSS} emerged as the most successful
candidate to describe the eleven dimensional theory.
Although it has already passed many nontrivial
tests, there remains nontrivial issues which
needs careful examinations.  One of such issues is
the Lorentz invariance.  Because of its very definition,
Matrix theory needs the extra information to understand
eleventh dimension (so called ``M''-direction).
Although there are some beautiful works \cite{r:PP} which
suggest the symmetry  by using
2+1 dimensional instanton calculus,
it is still desirable to have a direct confirmation.

The situation is essentially different in its close cousin,
de Wit-Hoppe-Nicolai (dWHN) supermembrane\cite{r:dWHN}.\footnote{
For a detailed information on  the supermembrane theory,
see \cite{r:TD} and references therein.}
Although the difference between the two theories is simply in
their gauge groups (SU(N) vs the area preserving diffeomorphism (APD)),
we have an explicit definition of the Lorentz generators \cite{r:dWMN}
and the Lorentz algebra itself was already checked explicitly
\cite{r:Mel}\cite{r:EMM}. 

In this letter, we examine the supermembrane in the
toroidally compactified spacetime.  
In section two, we propose Lorentz invariant form of the
SUSY algebra with the central charges associated with
membranes.
In section three, we derive the APD constraints
associated with the harmonic vector fields 
which play a central role in the analysis of the BPS conditions.
In sections four and five, we give equations that characterize 
BPS states with $1/2$ and $1/4$ SUSY.
Examination of the latter gives a system of 
the first order differential
equations which is analogous to the Bogomol'nyi bound
of the super Yang-Mills
theory. We show that a particular solution gives
the BPS states of the type IIA superstring after
the double dimensional reduction.
Finally in section six we discuss how our results may be
extended to the matrix formulation of M-theory.

\section{Eleven Dimensional SUSY algebra 
of Su\-per\-membrane and BPS condition}


Let us first examine the SUSY algebra of dWHN model.
We use the same notations and definitions
as in \cite{r:dWMN} in the following computation.
In particular the expression of supercharges is given by:
\ba
Q^{+}& = &
\frac{1}{\sqrt{P_{0}^{+}}}\int d^{2}\sigma
\left(P^{a}\gamma_{a}+\frac{\sqrt{w}}{2}\{X^{a},X^{b}\}
\gamma_{ab}\right)\theta,
\non
Q^{-}& = &
\sqrt{P_{0}^{+}}\int d^{2}\sigma\sqrt{w}\theta,
\ea
where $\{A,B\}\equiv\frac{\epsilon^{rs}}{\sqrt{w}}
\partial_{r}A\partial_{s}B$ $(r,s=1,2)$. Using the Dirac brackets:
$$
\left(X^{a}(\sigma),P^{b}(\rho)\right)_{DB}
=\delta^{ab}\delta^{(2)}(\sigma,\rho), \qquad
\left(\theta_{\alpha}(\sigma),\theta_{\beta}(\rho)\right)_{DB}=
-\frac{i}{\sqrt{w(\sigma)}}\delta_{\alpha\beta}\delta^{(2)}
(\sigma,\rho),
$$
the SUSY algebra of dWHN model is computed as follows
\cite{r:dWHN}(see also \cite{r:BSS}),
\ba
i\left(Q_\alpha^-, Q_\beta^-\right)_{DB}
 & = & \delta_{\alpha\beta}P_0^+,\non
i\left(Q_\alpha^-,Q_\beta^+\right)_{DB}
 & = & P_0^a(\gamma_a)_{\alpha\beta}
+\frac{1}{2} z^{ab}(\gamma_{ab})_{\alpha\beta},\non
i\left(Q_\alpha^+,Q_\beta^+\right)_{DB}
 & = & 2\delta_{\alpha\beta} H + 2z^a(\gamma_a)_{\alpha\beta}
+\frac{2}{4!}z^{abcd}(\gamma_{abcd})_{\alpha\beta}.
\label{e:SUSY}
\ea
The brane charges which appear in the right hand side
of these equations are defined by
\ba
z^{ab} & = & -\int d^2 \sigma \sqrt{w}\left\{
X^a, X^b\right\},\label{e:t2}\\
z^a & = & \frac{1}{P_0^+}\int d^2\sigma\left(
\left\{ X^a, X^b\right\}P_b - \frac{i}{2}\sqrt{w}
\left\{X^a,\theta^\alpha\right\}\theta^\alpha\right)
\non
& & -\frac{3i}{16P^+_0}\int d^2\sigma\sqrt{w}\left\{
X^c,\theta\gamma^{ac}\theta\right\},\label{e:l2}\\
z^{abcd} &=& -\frac{12}{P_0^+}\int d^2\sigma\sqrt{w}
\left\{X^{\left[a\right.},X^b\right\}\left\{
X^c,X^{\left.d\right]}\right\}-\frac{i}{4P^+_0}
\int d^2 \sigma \sqrt{w}\left\{ X^{\left[ a\right.},
\theta\gamma^{\left.bcd\right]}\theta\right\}\label{e:l4}.
\ea
The second term in \eq{e:l2}
and the second term in \eq{e:l4} should 
vanish as we already
discussed in our previous paper \cite{r:EMM} (appendix F)
to make the supercharge well-defined.
The first term in \eq{e:l4} vanishes for the membrane
configuration.  \eq{e:l4} should be 
regarded as the longitudinal
5 brane charge but it becomes 
absent in the supermembrane.
Finally, the first term in \eq{e:l2} can be rewritten as
\be
\int d^2 \sigma \sqrt{w} \left\{ X^-,X^a\right\}.
\ee
It makes the SUSY algebra \eq{e:SUSY}
Lorentz invariant\footnote{
We understand the SUSY algebra in this form
was also derived
by de Wit et. al \cite{r:dWPP}. 
We thank B. de Wit to send
us the preliminary version of their paper.
We have to admit that some part of this paper have
overlaps with theirs although it was studied
independently.}.


In \cite{r:BSS}, the BPS conditions of the SUSY algebra
was discussed in the Matrix theory.  
It is our purpose here 
to reexamine the analysis for
the manifestly Lorentz
invariant form \eq{e:SUSY}.
We write the SUSY algebra in the matrix form,
\ba
&&\left(
\begin{array}{cc}
i\left(Q^-,Q^-\right)_{DB} &
i\left(Q^-,Q^+\right)_{DB} \\
i\left(Q^+,Q^-\right)_{DB}&
i\left(Q^+,Q^+\right)_{DB} 
\end{array}
\right)  
=
\left(
\begin{array}{cc}
P_0^+ \cdot \id &  \sP+\szb\\
\sP-\szb & 2 H \cdot \id+2\sza
\end{array}
\right)
\non
&&\mbox{\hspace*{0.2in}}=
\left(
\begin{array}{cc}
P_0^+ \cdot \id & 0 \\
\sP-\szb & \id 
\end{array}
\right)
\cdot
\left(
\begin{array}{cc}
\frac{1}{P_0^+}\id & 0\\
0                  & \frac{1}{P_0^+}\cm
\end{array}
\right)
\cdot
\left(
\begin{array}{cc}
P_0^+ \cdot\id  & \sP+\szb \\
0          & \id
\end{array}
\right).
\label{e:scmat}
\ea
Here our notation is $\sP=P^a_0\gamma_a$,
$\sza=z^a\gamma_a$, $\szb = \frac{1}{2} z^{ab}\gamma_{ab}$.
The real symmetric matrix $\cm$ is defined as,
\ba
\cm &=& 2P^+_0(H\cdot\id+\sza)
-(\sP-\szb)(\sP+\szb)\non
& = & (2P^+_0\cdot H -P^a_0 P^a_0-\frac{1}{2}z^{ab}z^{ab})
\id + 2(P_0^+z^a - P_0^c z^{ca})\gamma_a\non
& & + \frac{1}{4}z^{ab}z^{cd}\gamma^{abcd}.
\label{e:cm}
\ea
{}From \eq{e:scmat} we find that $\cm$ is
positive semi-definite when the theory is quantized.

At this point, it is easy to observe that
the BPS condition of $1/2$ SUSY is simply $\cm=0$
and that of $1/4$ SUSY is that $\cm$ has rank 8.
We will analyze these conditions in detail in 
sections 4 and 5.

\section{Constraint from APD}

Associated with the gauge symmetry in the 0+1 dimensional
Yang-Mills system,
the Gauss law constraint of the dWHN model is given by
\ba
\varphi(\sigma) & = & -\left\{\frac{P^a}{\sqrt{w}}
, X^a\right\}-\frac{i}{2}\left\{ \theta,\theta\right\}
\approx 0\label{e:gauss1}\\
\varphi^{(\lambda)}
& = & \int d^2\sigma \epsilon^{rs}\phi^{(\lambda)}_r
\left( P^+_0\partial_s X^-+\frac{P^a}{\sqrt{w}}
\partial_s X^a +\frac{i}{2} \theta\partial_s\theta\right)
\approx 0.\label{e:gauss2}
\ea
The first constraint comes from the area preserving diffeomorphism
(APD) in the bulk.  The second ones are associated with
the harmonic one form $\phi^{(\lambda)}_r$ where $\lambda=1,\cdots,
2g$ ($g$ is the genus of the surface).  These two conditions
ensure the integrability of the definition of $X^-$,
\be
\partial_r X^-(\sigma) = -\frac{1}{P_0^+}
\left( \frac{P^a}{\sqrt{w}}\partial_r X^a + \frac{i}{2}
\theta\partial_r\theta\right).
\ee

When the target space has a toroidal topology,
\be
X^a\sim X^a+2\pi R^a, \quad X^-\sim X^-+2\pi R,
\ee
and the membrane
has certain winding number, the embedding coordinates and their
momenta can be expanded in terms of the eigenfunction of the
Laplacian as follows,\footnote{
We normalize the harmonic one forms $\phi^{(\lambda)}_{r}$ as
$$
\oint_{C^{\lambda^{\prime}}}d\sigma^{r}\phi^{(\lambda)}_{r}
=\delta^{\lambda\lambda^{\prime}},
$$
where $C^{\lambda}$ ($\lambda=1,2,\ldots,2g$) comprize a basis
of the first homology class.}
\ba
\partial_r X^a(\sigma)  & = & 2\pi R^a\phi^{(\lambda)}_r n^{(\lambda)a} + 
\sum_A X^a_A\partial _r Y^A(\sigma),\non
\partial_r X^-(\sigma)  & = & 2\pi R\phi^{(\lambda)}_r n^{(\lambda)} + 
\sum_A X^-_A\partial _r Y^A(\sigma),\non
P^a(\sigma) & = &P^{+}(\sigma)\frac{\partial}{\partial t}X^{a}(\sigma)
=\sqrt{w}\left(\frac{m^{a}}{R^a}+\sum_A  P_A^a
Y^A(\sigma)\right),
\non
P^+(\sigma) & = &\sqrt{w}P^+_0=\sqrt{w}\frac{m}{R},\non
\theta^\alpha(\sigma) & = & \theta_0^{\alpha}+
  \sum_A\theta^\alpha_A Y^A(\sigma).
\ea
Here $Y_A$ is the eigenfunction of the Laplacian with non-zero
eigenvalue, $\Delta Y_A = -\omega_A Y_A$, $\omega_A>0$.
$n^{(\lambda)a}$, $n^{(\lambda)}$, $m^{a}$ and $m$ are
integer-valued.
We plug the expansion into \eq{e:gauss2} to get
\ba
\varphi^{(\lambda)}& = &2\pi f_{\lambda\lambda'0}
(m n^{(\lambda')}+m^a n^{(\lambda')a})+2\pi
\sum_{\lambda',B}{ f_{\lambda\lambda'}}^{B}R^a n^{(\lambda')a}
P^a_B\non
&&+\sum_{AB}{f_{\lambda}}^{AB}(X^a_AP^a_B-\frac{i}{2}\theta_A\theta_B).
\ea
The structure constants are defined as
$$
f_{\lambda AB}=\int d^{2}\sigma\epsilon^{rs}\phi_{r}^{(\lambda)}
\partial_{s}Y_{A}Y_{B},\quad
f_{\lambda\lambda^{\prime}B}=\int d^{2}\sigma\epsilon^{rs}
\phi^{(\lambda)}_{r}\phi^{(\lambda^{\prime})}_{s}Y_{B}.
$$

In our analysis in the following sections,
we mainly take the topology of the membrane as two torus.
If we pick the coordinate $\sigma^r$ to satisfy $\sigma^r\sim
\sigma^r+1$ ($r=1,2$),
the eigenfunction becomes $Y^A = e^{2\pi i (A_1\sigma^1
+ A_2\sigma^2)}$ with $A=(A_1, A_2)\neq (0,0)$, $A_i\in\mbox{{\bf Z}}$.
We write the expansion of the embedding function as,
\ba
X^- & = & -\frac{R}{m} Ht + 2\pi Rn_{r}\sigma^{r}
+\hX^-(\sigma)\non
X^a & = & \frac{R}{m}\frac{m^a}{R^a} t
+2\pi R^a n^a_r\sigma^r+\hat{X^a}(\sigma),
\ea
where the periodic parts are given by
$
\hX^\mu(\sigma) = \sum_A X^\mu_A Y^A,
$
and so on.

The central charges of the toroidal membrane are given as
\ba
z^{ab} & = & -(2\pi)^2R^aR^b(n_1^an_2^b-n_2^an_1^b)\non
z^a & = & (2\pi)^2R R^a(n_1 n_2^a-n_2 n_1^a).
\ea
The constraint \eq{e:gauss2} is simplified as,
\be
2\pi\varphi_r \equiv -\epsilon_{rs}\varphi^{(s)}
= 2\pi\left( m n_r + m^a n^a_r +i
\sum_{A\neq \vec 0} A_r (P^a_{-A}X^a_A+\frac{i}{2}\theta^\alpha_{-A}
\theta^\alpha_A)\right)\approx 0.
\label{eq:levelmatching}
\ee
As we will see, this condition may be regarded as an analogue of
the level matching condition in string theory.

\section{BPS configuration with 1/2 SUSY}

In the following we will mainly consider
the case when the topology of the membrane
is two torus.  In such situation,
the last term in \eq{e:cm} vanishes which facilitates
the analysis of the BPS condition.

By using the definition of
the invariant supermembrane mass $\mass$:
\be
\mass^2=2P_0^+ \cdot H - P_0^a P_0^a,
\ee
the BPS condition $\cm = 0$ becomes\footnote{We point out that
$\mass^{2}$ is Lorentz invariant by virtue of $z^{a+}=0$.},
\ba
\mass^2 & = & \frac{1}{2}z^{ab}z^{ab}\non
P_0^+ z^a - P^c_0z^{ca} & = & 0.
\label{e:BPS}
\ea
The second condition relates the winding in the
longitudinal direction to those 
in the transverse dimensions.

%


Now we can discuss the relationship between (\ref{eq:levelmatching})
and the BPS condition \eq{e:BPS}. The first equation in \eq{e:BPS}
tells us that there is no nonzero mode contribution. 
This implies
\ba
X^a & = & \frac{R}{R^a}\frac{m^a}{m}t + 2\pi R^a
n^a_r\sigma^r ,\non
X^- & = & -\frac{R}{m}Ht+2\pi Rn_r\sigma_r ,\non
\theta^{\alpha} &=& \theta^{\alpha}_{0}.
\label{e:M-membrane}
\ea
The constraint \eq{eq:levelmatching} is reduced
 to the following simple relation
\be
\varphi_{r}^{(0)}\equiv mn_{r}+m^{a}n^{a}_{r}=0.
\ee
Using this, the second equation of \eq{e:BPS} is rewritten as
\be
\varphi_{1}^{(0)}n_{2}^{a}-\varphi_{2}^{(0)}n_{1}^{a}=0.
\ee
We therefore conclude that, by virtue of the constraint
\eq{eq:levelmatching}, the 1/2 BPS condition \eq{e:BPS}
is automatically satisfied even for the membrane
wrapping in the M-direction (with no nonzero modes).
This suggests that \eq{eq:levelmatching}
plays an important role in showing the Lorentz
invariance of the supermembrane theory.

%

\section{BPS configurations with 1/4 SUSY}

Let us proceed to explore the equation of supermembrane
with one quarter SUSY.
As before we assume the toroidal topology of the supermembrane.
The BPS condition is that
the matrix $\cm$ in \eq{e:cm} has rank 8.
Since $P_0^+ z^a - P^c_0z^{ca}$ is a constant vector,
we can always choose a nine-dimensional orthonormal basis
$(e^{(9)}_{a},e^{(i)}_{a})$ $(i=1,2,\ldots,8)$ with the property
\be
P_0^+ z^a - P^c_0z^{ca}\propto e^{(9)a}.
\ee
We will henceforth denote the  components of a vector $V^{a}$ as
\be
V^{9}=e^{(9)}_{a}V^{a},\quad
V^{i}=e^{(i)}_{a}V^{a}.
\ee 
In this frame, the BPS condition is equivalent to
\be\label{e:bps4}
\mass^2-\frac{1}{2} z^{ab}z^{ab}  \mp  2 
(P_0^+ z^9 - P^c_0z^{c9})=0.
\ee
We introduce the notation,
$\nabla^a \equiv 2\pi R^a(n^a_1 \partial_2 - n^a_2\partial_1)$.
Various parts in the Hamiltonian are written
as,
\ba
\left\{ X^a, X^b\right\} & = & -z^{ab} 
+ \nabla^a \hat{X}^b-\nabla^b\hat{X}^a+
\left\{ \hat{X}^a,\hat{X}^b\right\},\non
\left\{ X^a, P^b\right\} & = & 
\nabla^a\hat{P}^b + \left\{
\hat{X}^a,\hat{P}^b\right\}.
\ea
In the following analysis, we assume the fermionic background
to vanish for  simplicity.
The left hand side of \eq{e:bps4} becomes,
\ba
&&\int d^2\sigma \left(\hP^a\hP^a+\frac{1}{2}
(\nabla^a \hX^b - \nabla^b \hX^a+
\left\{\hX^a,\hX^b\right\})^2\right)
\mp 2\int d^2\sigma
\hP^c\nabla^9 \hX^c
\non
&=&\int d^2\sigma \left[
\left(\hP^c\mp(\nabla^9\hX^c-\nabla^c\hX^9
+\left\{ \hX^9,\hX^c\right\})\right)^2\right.\non
&&\qquad\quad\left.+\frac{1}{2}\left(\nabla^i\hX^j-\nabla^j \hX^i+ 
\left\{\hX^i,\hX^j\right\}\right)^2\right]\geq 0.
\ea 
In deriving this equation,
we used the APD constraints \eq{e:gauss1} \eq{e:gauss2}.
The final expression becomes a sum of the
squares as expected.
The BPS condition for $1/4$ SUSY becomes,
\ba
\hP^9 & = & 0 ,\non
\hP^i& = & \pm\left(\left\{ X^9, X^i\right\} +z^{9i}\right)  ,\non
0& = & \left\{X^i,X^j\right\} +z^{ij}.
\ea
This equation\footnote{Similar problem was
approached by Becker, Becker and
Strominger \cite{r:BBS} in a slightly different context.}
 is an analogue of the Bogomol'nyi bound
of the super Yang-Mills theory (see for example \cite{r:Hav}).
We note that, in the situation considered here, the following equation
holds
\be
P^+_i z^i-P^c_0 z^{ci} = 0.
\ee
The SUSY generators which are not broken under such a configuration are,
\ba
Q^{(\mp)} & \equiv & \Pi^{\mp}\left\{\sqrt{P^{+}_{0}}Q^{+}
-\frac{(\sP-\szb)}{\sqrt{P^{+}_{0}}}Q^{-}\right\}\non
& =  & \Pi^\mp \int d^2\sigma \left(
\hP^a \gamma^a + \frac{1}{2} \left(
\left\{ X^a,X^b\right\} + z^{ab} \right)
\gamma_{ab}\right) \htheta,
\ea
where $\Pi^\mp= (1\mp \gamma^9)/2$ are projection operators.
In fact these generators have vanishing Dirac brackets with canonical
variables, e.g.,
\ba
\left(Q^{(\mp)}, \theta\right)_{DB} & = & 
-i\Pi^\mp\left[ \mp \hP^9+\left(\hP^i\mp \left(
\left\{X^9, X^i\right\}+z^{9i}\right)\right)\gamma_j
+\frac{1}{2}\left(\left\{X^i,X^j\right\}+z^{ij}\right)\gamma_{ij}\right]
\non
&=&0.\label{e:Qtheta}
\ea
%
We remark that, in general, the right hand side  of \eq{e:Qtheta} need
not be strictly zero.
It is sufficient to set it zero modulo APD gauge transformations.
This enables us to analyze the case of non-vanishing $\theta$.

As an illustration let us consider the following configuration:
\begin{eqnarray}
X^{9}&=&\frac{R}{R^{9}}\frac{m^{9}}{m}t+2\pi R^{9}n^{9}\sigma^{2},
\nonumber \\
X^{i}&=&\frac{R}{R^{i}}\frac{m^{i}}{m}t+2\pi R^{i}n^{i}\sigma^{1}
+\hat{X}^{i}(t,\sigma^{1}),
\nonumber \\
\theta^{\alpha}&=&\theta^{\alpha}_{0}+\hat{\theta}^{\alpha}(t,\sigma^{1}),
\nonumber \\
X^{-}&=&-\frac{R}{m}Ht+2\pi R(n\sigma^{1}+n^{\prime}\sigma^{2})
+\hat{X}^{-}(t,\sigma^{1}),
\nonumber \\
P^{+}&=&\frac{m}{R}.
\end{eqnarray}
This configuration has the central charges
\begin{equation}
z^{9}=4\pi^{2}RR^{9}nn^{9},\quad
z^{i}=-4\pi^{2}RR^{i}n^{\prime}n^{i}, \quad
z^{9i}=4\pi^{2}R^{9}R^{i}n^{9}n^{i},\quad
z^{ij}=0.
\end{equation}
The Gauss law constraint for this configuration
reduces to\footnote{The constraint in the bulk,
$\varphi(\sigma)\approx 0$, is automatically satisfied in this case.
}
\begin{eqnarray}
\varphi_{1}&=&mn+m^{i}n^{i}+\frac{1}{2\pi}\int d\sigma^{1}
(\hat{P}^{i}\partial_{1}\hat{X}^{i}+\frac{i}{2}\hat{\theta}
\partial_{1}\hat{\theta})\approx 0,
\nonumber \\
\varphi_{2}&=&mn^{\prime}+m^{9}n^{9}\approx 0.
\end{eqnarray}
The first equation is of the same form as the level-matching
condition of the closed superstring.
This is consistent with the fact that, after the double dimensional
reduction, 11D supermembrane reduces to 10D type IIA superstring
\cite{r:DHIS}.
The BPS condition is rewritten as
\begin{eqnarray}
\partial_{t}\hat{X}^{i}&=&\mp2\pi RR^{9}\frac{n^{9}}{m}
\partial_{1}\hat{X}^{i},\nonumber \\
\Pi^{\pm}\hat{\theta}&=&0.
\label{eq:DDR}
\end{eqnarray}
The second equation comes from the condition:
$(Q^{(\mp)},X^{a})_{DB}=0$ {\em mod APD}.
It leads us to see that the fermion modes with
plus (minus) chirality are projected out.
Combined with equations of motion, the condition (\ref{eq:DDR})
picks up only the left(right)-handed modes
in the $\sigma^{1}$-direction. 
These configurations are therefore understood as an
extension of the BPS configurations in the type IIA
superstring to 11D supermembrane.

\section{Discussion}

In this paper we investigated winding modes
of the supermembrane in the light cone gauge.
We have obtained the following results: (i) 1/2 SUSY is
achieved even if the membrane wraps around the longitudinal direction;
(ii) success in constructing such configurations is attributed to
the Gauss law constraint \eq{e:gauss2} associated with the harmonic
vector fields; (iii) we derive the first order differential equations
to characterize $1/4$ SUSY; (iv) we explicitly constructed  string
BPS states from those of the membrane.

While the constraint \eq{e:gauss2} has been overlooked in the previous
analysis of  M(atrix) theory, it may  play an essential
role if our result is taken seriously.
Thus it may be useful to consider an extension of \eq{e:gauss2} to
M(atrix) theory. In the case of a toroidal supermembrane, we can
construct an obvious candidate:
\ba
\varphi^{(1)}_{M}=2\pi mn^{2}+{\rm Tr}\left([q,X^{a}]P^{a}
-\frac{i}{2}[q,\theta^{\alpha}]\theta^{\alpha}\right),\non
\varphi^{(2)}_{M}=-2\pi mn^{1}+{\rm Tr}\left([p,X^{a}]P^{a}
-\frac{i}{2}[p,\theta^{\alpha}]\theta^{\alpha}\right).
\label{eq:Mgauss2}
\ea
$X^{a}$, $P^{a}$ and $\theta^{\alpha}$ are now regarded as $m\times m$
matrix-valued and $(q,p)$ are the matrices
with the commutation relation $[q,p]=I$.
A candidate for the longitudinal membrane in M(atrix) theory is also
obtained if we replace $(\sigma^{1},\sigma^{2})$ in \eq{e:M-membrane}
by $(q,p)$.


One important point is that
the generalization of the $1/4$ condition to the M(atrix)
theory is straightforward.  All we have to do is to replace
the APD bracket with the commutator.

We hope that our approach gives a new viewpoint to this 
famous problem
and the relation with the BPS membrane state
in the supergravity theory
\cite{r:DS} will be very interesting.

\vskip 3mm
\noindent {\bf Acknowledgement: }

We would like to thank H. Nicolai and B. de Wit for communication.
We are also obliged to M. Ninomiya for encouragement.

\vskip 3mm



\begin{thebibliography}{9}

\bibitem{r:BFSS}
T. Banks, W. Fischler, S.H. Shenker and L. Susskind,
Phys. Rev. D55 (1997), 5112-5128,
``M theory as a Matrix Model: A Conjecture'', hep-th/9610043.
\bibitem{r:PP}
J. Polchinski and P. Pouliot, ``Membrane Scattering with
M-Momentum Transfer'', hep-th/9704029;\\
T.Banks, W. Fischler, N. Seiberg and L. Susskind,
``Instantons, Scale Invariance and Lorentz Invariance in Matrix
Theory'', hep-th/9705190
\bibitem{r:dWHN}
B. de Wit, J. Hoppe and H. Nicolai, Nucl. Phys. B 305 [FS23]
(1988), 545-581.
\bibitem{r:TD}
P. K. Townsend, ``Three lectures on supermembranes'',
in {\it Trieste School 1988: Superstrings: 438} ;\\
M. J. Duff, ``Supermembranes'', hep-th/9611203.
\bibitem{r:dWMN}
B. de Wit, U. Marquard and H. Nicolai, Commun. Math. Phys.
128 (1990), 39-62.
\bibitem{r:Mel}
S. Melosch, Diploma Thesis (1990) written in German, unpublished.
\bibitem{r:EMM}
K. Ezawa, Y. Matsuo and K. Murakami,
``Lorentz Symmetry of Supermembrane in Light Cone Gauge
Formulation'', hep-th/9705005, to appear in Prog. Theor. Phys..
\bibitem{r:BSS}
T. Banks, N. Seiberg and S. Shenker, ``Branes from Matrices'',
hep-th/9612157.
\bibitem{r:dWPP}
B. de Wit, K. Peeters and J. C. Plefka, ``Supermembranes
with Winding'', hep-th/9705225.
\bibitem{r:BBS}
K. Becker, M. Becker and A. Strominger,
Nucl. Phys. B456 (1995) 130, hep-th/9507158.
\bibitem{r:Hav}
J. A. Harvey, ``Magnetic Monopoles, Duality, and Supersymmetry'',
hep-th/9603086 and references therein.
\bibitem{r:DHIS}
M. J. Duff, P. S. Howe, T. Inami and K. S. Stelle,
Phys. Lett. B191 (1987) 70.
\bibitem{r:DS}
M. J. Duff and K. S. Stell, Phys. Lett. B253 (1991) 113;\\
R. G\"{u}ven, Phys. Lett. B 276 (1992) 49.
\end{thebibliography}
\end{document}